# A *Mathematica* Program Using Isotropic Harmonic Oscillator Eigenstates to Model Alpha-Particle Tunneling Half Life Variability

Eugene F. Chaffin, Nicholas W. Gothard, and Joshua Tuttle

Bob Jones University, Greenville, SC 29614

## Abstract

We have developed a *Mathematica* program which uses three-dimensional, isotropic harmonic oscillator wavefunctions for the solutions interior to the nucleus, and Coulomb wavefunctions for the exterior. The algorithm enables us to calculate the half life for alpha decay for varying depths of the nuclear potential well.

## Introduction

In recent years string theory the dependency of various constants on radii of compactified dimensions has made it appear that half-lives for alpha decay may have been variable during the early history of the universe [1-5]. In order to model this time dependence, we have used *Mathematica* to model the variation of the decay constant with change in depth of the potential well. More than 20 years ago, Pierronne and Marquez [6] treated the theory of alpha decay using a square well solution for the interior of the nucleus and coulombic solutions for the exterior, in a modification of the pioneer work by Preston [7]. The Pierronne and Marquez approach enables a finite well depth of around -50 to -80 MeV to be used, whereas the earlier approach of Preston required a unrealistic assumption of a positive value for the well depth. The use of the square well leads to spherical Bessel functions for the interior solutions of the square well, which Pierronne and Marquez matched to repulsive coulomb solutions on the boundary of the potential well. The spherical Bessel functions have the unfortunate disadvantage that the radial quantum number of the solutions is only defined implicitly.

To eliminate this problem, we have repeated the problem using the solutions for the isotropic harmonic oscillator in three dimensions.

## The Harmonic Oscillator Solutions

As is well known, the radial solutions for the harmonic oscillator can be written in terms of the confluent hypergeometric function:

$$F_{v\ell}(r) = \sqrt{\frac{2\lambda^{2\ell+3}\Gamma\left(v+\ell+\frac{3}{2}\right)}{v!\left[\Gamma\left(\ell+\frac{3}{2}\right)\right]^2}} r^\ell \, e^{-\frac{(\lambda r)^2}{2}} \, {}_1F_1\left(-v, \ell+\frac{3}{2}; (\lambda r)^2\right) \qquad (1)$$

where $\lambda$ is the oscillator constant, which is assumed to be a complex number in this approach. As in the Pierronne and Marquez, we match the logarithmic derivatives of the $\ell = 0$ exterior solutions with those of the interior solutions:

$$\frac{1}{rF_{v0}}\frac{d(rF_{v0})}{dr} = \frac{1}{F_{v0}}\frac{dF_{v0}}{dr} + \frac{1}{r} \qquad (2)$$

Ordinarily, the harmonic oscillator potential is zero at the origin and climbs to infinity for large radii. We choose instead to add a constant U to the potential, which does not change the form of the solutions, but enables the potential to be negative for small enough values of the radius r.

The derivatives of the $F_{v\ell}$ radial solutions are found rather easily using ladder operator relationships of these functions. The energy eigenvalue $E = \hbar\omega(2v + \ell + 3/2) + U$, where $\lambda^2 = m\omega/\hbar$, is required to have a small imaginary part, giving an energy $E - i\Gamma/2$, where $\Gamma$ is the decay width. This means that $\lambda^2 = \lambda_x^2 + i\lambda_y^2$ is a complex number, so that both the real and imaginary parts of the logarithmic derivative are matched to give the solutions. For the purpose of the calculation, we expand both sides of the equations, assuming $\lambda_y^2$ is small relative to $\lambda_x^2$, and this leads to simplified results, similar to the square well case.

## Description of the Algorithm

The algorithm requires input giving the atomic number Z, the atomic mass number A, the kinetic energy of the emitted alpha particle $E_\alpha$ and the depth U of the potential well. The output then gives the corresponding decay constant.